# Relic High Frequency Gravitational waves from the Big Bang and How to Detect Them

Andrew Beckwith

*abeckwith@uh.edu*

**Abstract.** This paper shows how entropy generation from numerical density calculations of relic gravitons can be measured via the Li-Baker high-frequency gravity wave (HFGW) detector, and suggests the implications this has for the physics of early-universe phase transitions. This paper indicates the role of Ng's revised statistics in gravitational wave physics detection and the application of Baumann *et al*. (2007) formalism of reduction of rank-two tensorial contributions to density wave physics, using the HFGW approximation directly at the beginning as well as Li's treatment of energy density explicitly. This formalism is a way to refine and add more capacity to the Li-Baker HFGW detector in reconstructing early-universe conditions at the onset of the big bang. Furthermore, we bring up how the HFGW detector can have its data sets compared and swapped with ice cube relic neutrino physics data taken at the south pole. This will enable us to begin to get criteria to falsify different inflation models as alluded to at the end of this manuscript.

## INTRODUCTION

At the June 2008 Dark Side meeting in Cairo, Egypt, Ng (2007) presented cogent arguments that entropy density is proportional to the number of dark matter particles per unit volume, which could also apply to gravitons. Ng's central idea—that entropy and numerical production of "particles" can be applied to density of created gravitons per unit volume—is based on an argument Weinberg (1972) used to calculate the number of gravitons per unit volume in a frequency range between $\omega, \omega + d\omega$. Weinberg's conceptualization of the creation of relic gravitons permits the development of a model entropy growth, starting from a very low level at the beginning of the universe to a much higher level right after the onset of the big bang, where the upper limit for the frequency used in deriving graviton production per unit volume was given by Grishchuk (2007) as $10^{10} Hz$, with some variance. This new model:

1) Assumes a temperature of $T \sim 10^{32} K$, based on Weinberg's (1972) statement that $T \sim 10^{32} K$ is the threshold for when quantum gravity dominates classical gravity,
2) Uses high-frequency gravitational waves (HFGW) explicitly of a scalar field reduction of the rank-two tensor arguments, based on the Baumann *et al*. (2007) use of quantum gravity operators to reduce a rank-two argument to a scalar field and then transform the entire object to a momentum space to get a scalar value for the variation of $g_{uv}$ due to gravity in momentum space, and
3) Uses Fourier analysis to extract relic gravitational wave signatures of the big bang, which is related to CMBR physics.

Ng (2007) established a one-to-one relationship between change in entropy and change in the number of particles. Equating entropy change and graviton particle production suggests that the Li-Baker detector (Li *et al*., 2008) could be used to get an explicit big bang signature from HFGW data sets. That is, the Li-Baker detection methods can be used as a basis of falsifiable experimental criteria for the existence of relic gravitons from the big bang.

The Li-Baker detector (Li *et al.*, 2008) uses a static magnetic field that varies under the impact of HFGWs for background detection of gravitational waves, and a fractal membrane to detect HFGW electromagnetic signatures. Researchers can use these signatures to confirm the existence of HFGWs by assuming that spin entropy density in the Li-Baker detector affects magnetic field spin and magnetism, per Rothman and Boughn (2006). Measurement of gravitons and gravitational waves is a way to establish an association between relic gravitational waves, gravitons, and entropy. This is a follow-up to a suggestion made by Yeo, *et al.* (2006) for how variation of spin entropy in a detector could dramatically enhance the sensitivity of existing HFGW detectors.

The benefit of examining spin entropy density is in showing the existence of gravitons as a physically measurable datum in General Relativity, as well as the interrelationship of gravitons with HFGW from experimental data sets. Rothman and Boughn (2006) have written that the present set of existing detector systems with pre-Li-Baker detector technology (Li *et al.*, 2008) is insufficient to accomplish meaningful detection of gravitons, suggesting that the Li-Baker detector may overcome the limits described by Rothman and Boughn: that with conventional detectors, one would need a detector mass about the size of Jupiter to detect a single graviton.

Finally,we can compare how this re formulation may allow for data set swapping of information between the HFGW graviton production data and ice cube, which may improve our understanding of CMBR issues.

## REVIEWING JACK NG'S ARGUMENTS: HOW ENTROPY IS PROPORTIONAL TO A NUMERICAL DENSITY VALUE

The fact that in both the dark matter and in the relic graviton production cases, entropy has similar quantum Boltzmann statistics will be the starting point for a derivation of the production of relic gravitons, linked to falsifiable experimental measurements. Ng (2007) used the following approximation for temperature and its variation with respect to a spatial parameter, starting with temperature $T \approx R_H^{-1}$, where $R_H$ can be thought of as a spatial representation of a region of space in which one can acquire statistics for the particles in question. Assume that the volume of space to be analyzed is of the form $V \approx R_H^3$. Then look at a preliminary numerical factor. Here the proportionality argument of $N \sim (R_H/l_P)^2$ is made, where $l_P$ is the Planck's length $(\sim 10^{-35} cm)$ and a "wavelength" parameter $\lambda \approx T^{-1}$ is also specified. That is, the value of $\lambda \approx T^{-1}$ and of $R_H$ are approximately within an order of magnitude of each other.

Ng (2007) changes conventional statistics by outlining how to get $S \approx N$, which with additional arguments is defined to be $S \approx <n>$, the numerical density of a species of particles. Ng begins with a partition function

$$Z_N \sim \left(\frac{1}{N!}\right) \cdot \left(\frac{V}{\lambda^3}\right)^N, \qquad (1)$$

which according to Ng, leads to a limiting value of entropy of

$$S \approx N \cdot \left(\log\left[V/N\lambda^3\right] + 5/2\right), \qquad (2)$$

but with $V \approx R_H^3 \approx \lambda^3$. If $N$ is greater than one, entropy in equation (2) has a negative value. For a quantum Boltzmann statistic calculation to obtain entropy, one does not want entropy with a negative value. The positive valued nature of entropy for physical systems calculated by Boltzmann statistics is a convention of statistical physics. Now this is where Ng introduces the removing of the $N!$ term in equation (1). Inside the log expression, the expression of $N$ in equation (2) is removed. This is a way to obtain what Ng refers to as Quantum Boltzmann statistics, where for a sufficiently large $N$,

$$S \approx N. \qquad (3)$$

The supposition here is that the value of $N$ is proportional to the numerical graviton density $\langle n \rangle$. It is noted that equation (3) gives credence not only to Baker *et al.* (2008) being applied to gravitons, but the same effort as done by Li *et al.* (2007) and as proposed (Beckwith 2008) in a symposium at Chongqing University in October 2008, for astrophysical applications of gravitational waves. Sensitive applications of equation (3) will help confirm the breakthrough physics of how gravitons disturb uniform magnetic fields within a HFGW detector, as remarked by Li *et al.* (2006).]

# WEINBERG'S 1972 NUMERICAL ESTIMATE: THE NUMBER OF GRAVITONS PER FREQUENCY RANGE

Assuming that $\bar{k} = 1.38 \times 10^{-16} erg/{}^0K$, where ${}^0K$ denotes Kelvin temperatures and where gravitons have two independent polarization states, the number of gravitons per unit volume with frequencies between $\omega$ and $\omega + d\omega$ is given by Weinberg (1972) as

$$n(\omega)d\omega = \frac{\omega^2 d\omega}{\pi^2} \cdot \left[ \exp\left(\frac{2 \cdot \pi \cdot \hbar \cdot \omega}{\bar{k}T}\right) - 1 \right]^{-1} \tag{4}$$

The hypothesis presented here is that input thermal energy (from the prior universe) inputted into an initial cavity/region (dominated by an initially configured low temperature axion domain wall) would be thermally excited to reach the regime of temperature excitation. This would permit an order of magnitude drop of axion density $\rho_a$ from an initial temperature $T_{dS}\big|_{t \leq t_P} \sim H_0 \approx 10^{-33} eV$. [Per Beckwith (2008), this calculation assumes that $E_{graviton} \equiv \hbar \omega_{graviton} \propto (volume) \cdot \left[ energy\ density \equiv t_0^0 \right]$, where the energy density term is from GR formulas.

# GIVING FREQUENCY/ ENERGY VALUE INPUTS TO GRAVITONS FROM GR ENERGY DENSITY EQUATIONS

the Li *et al.* (2008) derivation/formula for energy density of gravitational waves is

$$t_0^0 \equiv \frac{c^4 k^2}{4\pi G a^3} \cdot \left[ h_\oplus^2 + h_\otimes^2 \right], \tag{5}$$

where $a \approx a_{initial} \cdot \exp(H_{initial}\tau)$, where $H_{initial}$ is the initial value of the Hubble expansion parameter, and $\tau$ is a conformal time value. This value for an exponentially expanding scale factor will be crucially important in what is calculated later.

### The polarization values of relic gravitational waves

Let us now consider how to get appropriate $h_\oplus, h_\otimes$ values by using Baumann *et al.* (2007) very complete treatment of rank-two tensorial contributions to the evolution of the gravitational wave contributions to entropy. This will be helped by having HFGW as a template to simplify a search for appropriate $h_{ij}$ behavior, which will be simplified after the reduction of $h_{ij}$ to a scalar field value. The main centerpiece of the derivation is to take into account a right-hand-side contribution of stress and strain to the conformal time evolution of $h_{ij}$, which in a scalar Baumann, *et al.* (2007) field contribution reduction of complexity, and leads to the fast Fourier transform (FFT) $\Im h_{ij} \equiv \hat{h}$ with an equation in conformal time $\tau$ that can be written as

$$\hat{h} \equiv \frac{A_1(k)}{a(\tau)} \exp\left(i\left[\vec{k} \cdot \vec{x} - k\tau\right]\right) + \frac{A_2(k)}{a(\tau)} \exp\left(i\left[\vec{k} \cdot \vec{x} + k\tau\right]\right) \tag{6}$$

As Li *et al.* (2008) writes, the expression for $\hat{h}$ in equation (6) is in response to a metric is written as

$$g_{\mu\nu} \equiv \begin{pmatrix} -a^2 & 0 & 0 & 0 \\ 0 & a^2(1+h_\oplus) & a^2 h_\otimes & 0 \\ 0 & a^2 h_\otimes & a^2(1-h_\oplus) & 0 \\ 0 & 0 & 0 & a^2 \end{pmatrix} \tag{7}$$

Changes in the treatment of equation (6) have to be made in order to consider a scalar expansion at the onset of the big bang, which would entail looking at stress and strain contributions to the evolution of the scalar field contribution to gravitational radiation, starting at the onset of the big bang. This treatment of space-time geodesics is modified after stress and strain processes are added to the evolution of the gravitational waves. Addition of stress and strain as presented by Baumann *et al.* (2007) leads to the following evolution equation of space-time deformations and gravitational wave evolution, where pressure $p$ is a constant and $T^i_j$ is a stress term. Furthermore, $k^2 \propto$ energy and $|a''/a| \propto$ potential energy so that

$$\hat{h}'' + 2\frac{a'}{a}\cdot\hat{h}' + k^2\hat{h} = 16\pi\cdot G\cdot a^2\cdot\left[\Pi_k(\tau) = \Im\left(T^i_j - p\delta^i_j\right)\right]. \tag{8}$$

Numerous Bessel and Hankel equations, referenced in Arfken (1985), show how combined solutions of Bessel and/or Hankel equations solve the homogeneous part of equation (9) above, provided that $\Pi_k(\tau) = 0$. If one wishes to take into account stress and strain forces associated with the onset of the big bang, one would have to look at particular and general solutions that use combinations of equations (8) and (9). The solution to equation (8) is based on what Baumann *et al.* (2007) developed in 2007 to deal with relic inflationary contributions to gravitational waves. The particular solution of equation (8) above will involve a Greens function treatment of equation (9), as described below, as an integral solution for $h$.

Typically, as seen in Arfken (1985), this means putting a delta function on the right-hand side of equation (8), and using the resulting solution of equation (8) as modified, times the right-hand side of equation (8) as the integrand for a particular solution, then integrating over conformal time $\tau$. In this situation, the very convenient $a(\tau)\cdot\hat{h} = \mu(k)$ is taken advantage of to use the Greens function solution to equation (9) with a delta function on the right-hand-side of equation (9) to help construct a particular solution to equation (8). This then will be part of how a particular solution for gravitational wave amplitude evolves in space-time.

## HFGW in Relic Inflationary Conditions

Now the homogeneous and particular solution for equation (8) above is looked at with comments on HFGW modifications that simplify matters enormously. This will be pertinent to Li, *et al.* (2008) and what will be discussed later in this paper about the Li-Baker HFGW detector system, with its uniform magnetic field impinged upon by incident HFGW. This leads to a experimentally falsifiable claim that before the onset of the CMBR formation 280,000 to 300,000 years after the big bang, data sets are signatures of phase transitions that modeled appropriately with the following formalism.

After making the substitution of $a(\tau)\cdot\hat{h} = \mu(k)$, equation (8) leads to a non-homogeneous perturbed Schrodinger-like equation, which can be written as

$$\mu'' + \left(k^2 - \frac{a''}{a}\right)\mu = a\cdot\left[16\pi\cdot a^2\cdot\Pi_k(\tau)\right], \tag{9}$$

where the particular solution to equation (8) becomes

$$\hat{h}_{Particular} = \frac{1}{a(\tau)}\cdot\int d\tilde{\tau}\cdot g_k(\tau,\tilde{\tau})\cdot\left(16\pi\cdot G\cdot\Pi_k(\tilde{\tau})\right) \tag{10}$$

The kernel in equation (10), $g_k(\tau,\tilde{\tau})$, obeys the following equation if $a''/a \ll k^2$, so for

$$g_k'' + \left(k^2 - \frac{a''}{a}\right)\cdot g_k \equiv \delta(\tau - \tilde{\tau}), \tag{11}$$

where

$$g_k(\tau,\tilde{\tau}) = \frac{1}{k}\left[\sin(k\tau)\cdot\cos(k\tilde{\tau}) - \sin(k\tilde{\tau})\cdot\cos(k\tau)\right]. \tag{12}$$

Given the above, a particular solution may be written as

$$\hat{h}_{Particular} = \frac{1}{a(\tau)k}\cdot\int d\tilde{\tau}\cdot[\sin(k\tau)\cos(k\tilde{\tau}) - \sin(k\tilde{\tau})\cos(k\tau)]\cdot\left(16\pi\cdot G\cdot\Pi_k(\tilde{\tau})\right) \tag{13}$$

Details for the $16\pi\cdot G\cdot\Pi_k(\tilde{\tau})$ part of this particular solution will be presented in the next section; for now, the general solution is presented. The main dynamics of the $16\pi\cdot G\cdot\Pi_k(\tilde{\tau})$ terms are that they are in part linked to quantum fluctuation. That is, the stress and strain are initially nucleated from a vacuum template of space-time itself in the beginning of a new universe, allowing for the following homogeneous part of evolution equation (8), with $\Pi_k(\tau) = 0$, where the homogeneous solution to equation (8) is based on Izquierdo (2006)

$$\hat{h}'' + 2\frac{a'}{a} \cdot \hat{h}' + k^2 \hat{h} = 0. \tag{14}$$

In the initial phases of nucleation of a new universe, equation (15) can be simplified as

$$\hat{h}'' + 2H_{initial} \cdot \hat{h}' + k^2 \hat{h} = 0. \tag{15}$$

Traditional treatments of both equations (14) and (15) make use of a dynamical, changing value of $a'/a$, in many cases leading to Bessel/Hankel equation solutions. By setting $a'/a \sim H_{Initial}$, to obtain

$$\hat{h}_{Total} = \hat{h}_{Initial-Value} \cdot \left[\exp(-H_{Initial}\tau)\right] \cdot \cos(k\tau + c_1) + \hat{h}_{Particular} \tag{16}$$

implies that in later times the dynamics are largely dominated by the particular, specialized solution.

## Stress and strain contributions to space time due to early universe production of HFGW

The following analysis will deal with the HFGW contribution to forming the $16\pi \cdot G \cdot \Pi_k(\tilde{\tau})$ stress and strain contribution, using much of what Baumann *et al.* (2007) set for the simplest case of how to evaluate $16\pi \cdot G \cdot \Pi_k(\tilde{\tau})$. This takes into account a simplified treatment of the Bardeen and Wagoner (1971) potential for times $\tau < \tau_{Threshold}$; effectively confining $\tau < \tau_{Threshold}$ to within with two orders of magnitude of the Planck's time interval after big bang nucleation of the present universe.

This means working with the following template for the stress-strain-vacuum nucleation problem

$$\left(16\pi \cdot G \cdot \Pi_k(\tilde{\tau})\right) \equiv S_k(source) = \int d^3\tilde{k} \cdot e(k,\tilde{k}) \cdot f(k,\tilde{k},\tau) \cdot \psi_{k-\tilde{k}} \cdot \psi_{\tilde{k}} \tag{17}$$

where $\psi_{\tilde{k}}$ is a quantum fluctuation (offering a simplified model) and the term $e(k,\tilde{k})$ is equal to $\tilde{k}^2 \cdot \left(1 - (\vec{k} \cdot \vec{\tilde{k}})/k\tilde{k}\right)$. The main result of this section will be to present $f(k,\tilde{k},\tau)$, where $w \propto 1/3$ is used to obtain

$$f(k,\tilde{k},\tau) \equiv \frac{4}{3 \cdot (1+w)} \left\{ \begin{array}{l} \dfrac{2 \cdot (5+3w)}{\left(1+\left|k-\tilde{k}\right|^2 \tau^2\right)} \cdot \dfrac{1}{\left(1+\left|\tilde{k}\right|^2 \tau^2\right)} \\ +4 \cdot \left[\dfrac{2\tau}{1+\left|k-\tilde{k}\right|^2 \tau^2} + \tau^2 \cdot \dfrac{\partial}{\partial \tau}\left(\dfrac{1}{1+\left|k-\tilde{k}\right|^2 \tau^2}\right)\right] \cdot \dfrac{\partial}{\partial \tau}\left(\dfrac{1}{1+\left|\tilde{k}\right|^2 \tau^2}\right) \end{array} \right\} \tag{18}$$

Note that this uses the Bardeen and Wagoner (1971) potential in early times, which is

$$\Phi = \frac{1}{1+k^2\tau^2} \tag{19}$$

[Note that the derived quantity of $\hat{h}$, which is a FFT, with quantum raising and lowering operator considerations added, will require an inverse FFT used in the $h_\oplus^2 + h_\otimes^2$ expression of $t_0^0$ for the Li-Baker detector.]

## A Simplified Quantum Fluctuation Model

Here, the ideas of Mukhanov and Wintizki (2007) are used, where they give a quantum fluctuation in $k$ space along the lines of:

$$\psi_k'' + (k^2 + m^2)\psi_k \cong 0 \tag{20}$$

In the limit of low mass, this will lead to

$$\psi_k \sim \exp(ik\tau) \tag{21}$$

The assumption is made that with additional data acquisition, the nucleation quantum fluctuation formula as outlined in equation (21) will be given considerably more structure.

## TIES TO THE LI-BAKER HFGW DETECTOR

With reference to Beckwith (2008), a power law relationship, first presented by Fontana (2005) using Park's (1955) earlier derivation, is presented as

$$P(power) = 2 \cdot \frac{m_{graviton}^2 \cdot \hat{L}^4 \cdot \omega_{net}^6}{45 \cdot (c^5 \cdot G)} \tag{22}$$

With the effective energy $E_{eff} \equiv \langle n(\omega) \rangle \cdot \omega \equiv \omega_{eff}$, where graviton production is connected to equation (22), with $\omega_{eff} \approx \omega_{net}$ or $\omega_{net} \to \omega_{eff}$.

This expression in power should be compared with the one presented by Giovannini (2008), averaging the energy-momentum pseudo tensor to get his version of a gravitational power energy density expression

$$\bar{\rho}_{GW}^{(3)}(\tau, \tau_0) \cong \frac{27}{256 \cdot \pi^2} H^2 \cdot \left(\frac{H}{M}\right)^2 \cdot \left[1 + \vartheta \cdot \left(\frac{H^4}{M^4}\right)\right] \tag{23}$$

This led Giovannini to state that "should the mass scale be picked such that $M \sim m_{Planck} \gg m_{graviton}$, and if the above formula were true, there are doubts that there could be inflation."

It is clear that gravitational wave density is faint, even if one makes the approximation that $H \equiv \dot{a}/a \cong m\phi/\sqrt{6}$ as stated by Linde (2008). So $H \equiv \dot{a}/a \cong m\phi/\sqrt{6}$ and $\dot{\phi} = -m\sqrt{2/3}$ makes it appropriate to use different procedures to come up with relic gravitational wave detection schemes to get quantifiable experimental measurements of relic gravitational waves. Equation (22), and Equation (24) below imply a very short lived, but extreme energy flux as the starting point for graviton production, and also implies a large requisite value for $M = V^{1/4}$, where $V$ would be unusually large as a requisite inflationary potential. This is to be expected if there is an input from a universe prior to our own, as outlined by Beckwith (2008a and 2008b). It also is related to the following argument as given by Linde (2008) for gravitational wave amplitude.

If one makes use of the present day gravitational radiation as $M = V^{1/4}$ (Kofman, 2008), the energy scale with potential $V$ and frequency

$$f \cong \frac{(M = V^{1/4})}{10^7 GeV} Hz, \tag{24}$$

implies that $f \sim 10^{10} Hz$. However, using equation (24) assumes that the temperature of thermally induced vacuum energy is rising to a maximum value $5T^* \approx 10^{32}\ ^0K$, which is a huge energy flux.

Beckwith (2008) asserted that midway during the thermal/vacuum energy transfer from a prior to the present universe, a relic graviton burst would have occurred. As shown in Table 1, this is consistent with a wormhole introduction of vacuum energy from a prior universe to the present, with a thermal buildup from near-zero vacuum energy values. The threshold burst is then consistent with a buildup of temperature from a prior universe, which introduces a relic graviton energy burst.

| TABLE 1. Graviton burst | | |
|---|---|---|
| Numerical values of graviton production | Temp | Scaled Power values |
| $N1 = 1.794 \times 10^{-6}$ | $T^*$ | 0 |
| $N2 = 1.133 \times 10^{-4}$ | $2T^*$ | 0 |
| $N3 = 7.872 \times 10^{21}$ | $3T^*$ | $1.058 \times 10^{16}$ |
| $N4 = 3.612 \times 10^{16}$ | $4T^*$ | ~1 |
| $N5 = 4.205 \times 10^{-3}$ | $5T^*$ | 0 |

By way of explanation (Beckwith, 2008), the above table assumes a rapid buildup of temperature resulting from energy-matter transfer from a prior universe. The Wheeler-De Witt wormhole equation, as given by Crowell (2005), contains a pseudo time component. The wormhole model of energy transfer uses Crowell's treatment of the Wheeler-De Witt equation to model a bridge from a prior universe to our present universe. At the time the temperature reaches a maximum value of $5T^*$ ($10^{32}$ degrees Kelvin), a graviton burst has already happened within $10^{-35}$ seconds, and the frequency has gone up to $10^{10}$ Hz, as given by

$$\Omega_{gw}(v) = \frac{\pi^2}{3} h^2(v) \left(\frac{v}{v_H}\right)^2 \tag{25}$$

in Grishchuk (2007) and charted in Figure 1.

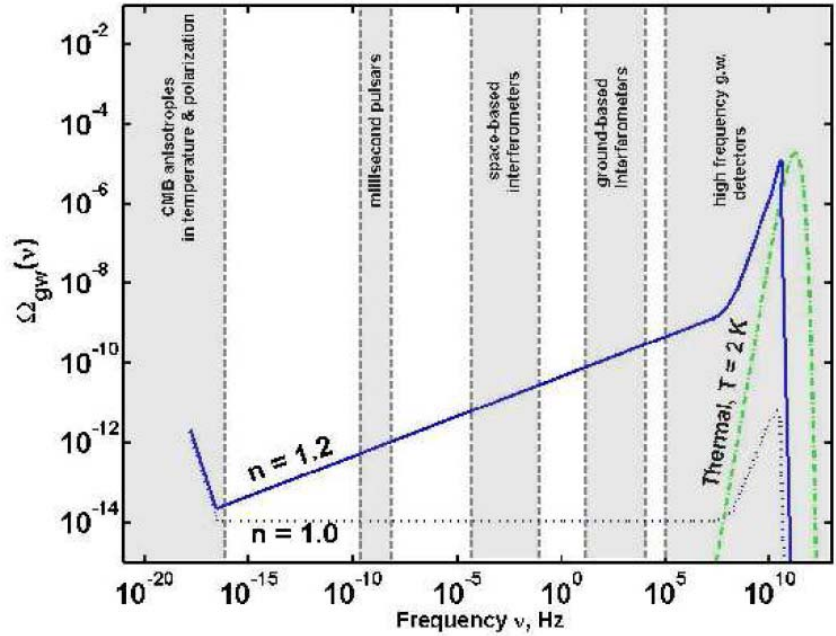

**FIGURE 1.** Where HFGWs come from: Grishchuk (2007) found the maximum energy density (at a peak frequency) of relic gravitational waves.

## Comparing results of numerical production

Smoot (2007) alluded to the following information theory regarding the number of information bits transferred between a prior and present universe:

1) Holographic principle allowed states in the evolution/development of the Universe - $10^{120}$
2) Initially available states given to us to work with at the onset of the inflationary era- $10^{10}$
3) Observable bits of information present due to quantum/statistical fluctuations - $10^{8}$

The relationship of bits to actual entropy, per Lloyd (2002), is that $10^{120}$ bits correspond to an entropy reading of $10^{90}$. Arguments by Carroll (2005) suggest that black holes in the center of galaxies have entropy readings of $10^{88}$, whereas the jump in entropy from about $10^{8}$ to $10^{90}$ is due to the jump in $<n>$, where $<n> \sim \Delta S_{graviton-production} \propto 10^{21}$.

In a meeting in Bad Honnef in April 2008, brane theorists suggested that the huge entropy reading of black holes in the center of galaxies is excusable, since most of the purported entropy would be hidden by the event horizon of black holes. From this and discussions with others, it is apparent that the event horizon of a black hole is equivalent to the escape velocity of a black hole, trapping huge amounts of information/entropy, since the escape velocity of a black hole is greater than the speed of light.

However, to measure entropy requires an entropy datum that can be measured. By definition, the black hole trapping of much of entropy in the universe leads to non-measurable data. Furthermore, it is assumed that the increase in entropy due to identification of a change of a relic graviton particle $<n> \sim \Delta S_{graviton-production} \propto 10^{21}$ is within the ability of the Li- Baker HFGW detector (Li *et al*., 2008) to obtain data sets.

Identifying this relic graviton burst would allow for understanding how entropy increased in the first place and would permit astrophysicists to model different phase transitions in the problem of how the universe traversed through the "graceful exit problem" in inflationary cosmology. Gasperini *et al.* (1996) modeled graceful exit from inflation in terms of the Wheeler-De Witt equation and phase transitions. So far, little new progress has been made in getting the data sets needed to ascertain if their suggestion is true, but the Li-Baker detector should be able to identify falsifiable data collection procedures to confirm or falsify Gaperini's suggestion of graceful exit from inflation.

## CONTRIBUTION OF THE LI-BAKER DETECTOR

According to Li et al. (2008), the Li-Baker detector (Li, *et al*., 2008) is able to measure the interaction of HFGWs with a static magnetic field $\hat{B}_y^{(0)}$, allowing researchers to get data on relic HFGWs created from relic big bang conditions. The electric and magnetic fields are generated by the HFGWs which, when $A_\otimes \approx A_\oplus \approx A(k_g)/a(t)$, and $\omega_g \leq 10^{10} Hz$ give the following values of electric and magnetic fields:

$$\tilde{E}_x^1 = \frac{i}{2} A_\oplus \hat{B}_y^{(0)} k_g c \cdot (z+l_1) \exp[i(k_g z - \omega_g t)] + \frac{1}{4} A_\oplus \hat{B}_y^{(0)} c \exp[i(k_g z + \omega_g t)]$$

$$\tilde{B}_y^{(1)} = \frac{i}{2} A_\oplus \hat{B}_y^{(0)} k_g (z+l_1) \exp[i(k_g z - \omega_g t)] - \frac{1}{4} A_\oplus \hat{B}_y^{(0)} \exp[i(k_g z + \omega_g t)]$$

$$\tilde{E}_y^{(1)} = -\frac{1}{2} A_\otimes \hat{B}_y^{(0)} k_g c \cdot (z+l_1) \exp[i(k_g z - \omega_g t)] + \frac{i}{4} A_\otimes \hat{B}_y^{(0)} \exp[i(k_g z + \omega_g t)]$$

$$\tilde{B}_z^{(1)} = \frac{1}{2} A_\otimes \hat{B}_y^{(0)} k_g (z+l_1) \exp[i(k_g z - \omega_g t)] + \frac{i}{4} A_\otimes \hat{B}_y^{(0)} \exp[i(k_g z + \omega_g t)]$$

(26)

a frequency value of $\omega_g \leq 10^{10} Hz$ is a base line for measurement in the Li- Baker detector in hopefully soon to be available data sets. Li *et al.,* (2008) numerically simulated incident relic graviton flux detected by the Li-Baker detector, with a value of $N_g \cong 2.89 \times 10^{14} / \sec$ at a detector site. Beckwith (2008) has also created a model that simulates graviton flux, for all gravitons produced by the big bang, of $<n>_g \sim 7.872 \times 10^{+21} / \sec$ (not just the gravitons detected by the Li-Baker detector)

## Quantum Entanglement

Based on the same numerical simulation done by Dr. Li and reported in , Li *et al.*, (2008), Dr. Li made a prediction (see equation 27) about the number of HFGW/ gravitons produced by the big bang, as compared to the general number of HFGW/ gravitons which the Li- Baker detector can access at the site of the detector. The difference in the numerator and denominator of equation (27) makes the case for the use of quantum entanglement detectors. There is a gap , i.e. a difference in the number of HFGW gravitons that are detectable by the Li-Baker detector (for all time from the big bang up to the present) and those relic HFGW gravitons from the onset of the big bang. The difference between gravitons producted at the onset of the big bang and those which are generally accessible for all times , from the big bang to the present is given in equation (27) below. The ratio of $10^{-2}$ appearing in the square root of equation (27) means that one out of a incoming HFGW gravitons detected by the Li-Baker detector would be relic in origin—directly due to the big bang—whereas the other 99 gravitons are due to astrophysical processes occurring *after* the big bang. The significance of the $8.76 \times 10^{-2}$ value in equation (27) below in Li, et al (2008) is that this number refers to the ratio of the strength of the amplitude of the different gravitational wave contributions. I.e. the amplitude of the gravitational waves from the big bang are $\sim 10^{-1}$ weaker at the Li Baker detector than general HFGW detected at the detector itself. What Li refers to as the strength, overall magnitude, of a PPF is the square root of numerical graviton flux, $\sqrt{N_g}$ . So equation (27) refers to the overall magnitude , amplitude difference in value in the process of graviton production in the origins of the universe, to what is seen today. The values of $10^{-1}$ and $10^{-2}$ come directly from Dr. Li et al(2008) discussions of what is their equation (62) in their article written in 2008.

$$\sqrt{\frac{N_{g-relic-GW}}{N_{g-plane-GW}}} \equiv \sqrt{\frac{2.89 \times 10^{14}}{3.77 \times 10^{16}}} \approx 8.76 \times 10^{-2} \tag{27}$$

This rarity of relic big bang gravitons means that for a next-generation refinement of sensitivity, the Li-Baker detector would need to use a variant of quantum entanglement to obtain a better data set of relic HFGW originating in the big bang. Yeo *et al.* (2006) presented calculations showing that a passing gravitational wave could influence the spin entropy and spin negativity of a system of *N* massive spin-(1/2) particles, in a way that is characteristic of the radiation. This implies entanglement, as Yeo *et al*. (2006) suggested. This suggests that what is now needed is to develop an actual entanglement entropy-based device that could complement and give additional refinement to the prediction given by equation (27). That would then help to analyze HFGWs to determine relic $\omega_{gravitation}$ values.

## Thoughts and commentary upon swapping HFGW data with relic neutrinos

As a convenience, this paper uses estimates from previous work (Beckwith, 2008) for high frequency gravitational wave (HFGW) upper frequency of detectable gravitational waves of approximately $10^{10} Hz$ . Furthermore, arguments for an upper limit for the mass of neutrinos will be reviewed, and a mean temperature of $T \sim 10^{32} K$ is assumed as a starting point for neutrino and graviton production.

Valev (2006) benchmarks upper limits for gravitational mass, as traditionally calculated, by claiming that observations lead to upper-bound values of

$$m_g \sim 1.2 \times 10^{-37} eV/c^2 \tag{28}$$

And

$$m_{\nu_{e^-}} \sim 0.0002 eV/c^2 \tag{29}$$

Furthermore, the ratio of about $10^5$ neutrinos to each graviton (most likely due to relic production of both ;neutrinos and gravitons) and the Li-Baker predictions (Li *et al.*, 2008; Baker, Stephenson and Li, 2008; Stephenson, 2009) for electric and magnetic field production suggests a connection between graviton/gravitational wave detection and neutrino data sets generated by the IceCube South Pole detector.

The key to comparing these two data sets is the the ratio of 1 graviton to 10 5th neutrinos. That is, the presence of very low, but existent masses for both of them, plus some specifics as to how the IceCubeIceCube detector works, as far as obtaining data sets will be employed to make the point. The HFGW planar wave approximation as given by Eqn (26) is due to first order perturbative photon flux (PPF). The PPF is generated by the interaction of HFGWs with a static magnetic field $\hat{B}_y^{(0)}$ (the inverse Gertsenshtein effect) and the synchro-resonance effect between the HFGWs and the background Gaussian beam. Because the transverse PPF (signal) and the background photon flux (BPF) have very different behaviors (e.g. propagating direction, distribution, phase, polarization, decay, etc.) in some special regions, such properties provide a new way to distinguish and display the perturbative effects of HFGWs. Li et al. (2008) and Baker, Stephenson and Li (2008) give the transverse PPF form as follows, with $\omega_e = \omega_g$ set as boundary conditions in the time-averaged brackets below. Equation (30) E and M fields are electromagnetic terms used to define $F_\alpha^{0(0)}$ and $\tilde{F}^{1\alpha(1)}$.

$$n^1 = \frac{c}{\hbar\omega_e} \cdot \left\langle \overset{(1)}{T}^{01} \right\rangle_{\omega_e=\omega_g} = -\frac{c}{\mu_0 \hbar\omega_e} \cdot \left\langle F_\alpha^{0(0)} \tilde{F}^{1\alpha(1)} + \tilde{F}_\alpha^{0(1)} F^{1\alpha(0)} \right\rangle_{\omega_e=\omega_g} \qquad (30)$$

The $F_\alpha^{0(0)}$ and the $\tilde{F}^{1\alpha(1)}$ terms represent the background and perturbative EM fields respectively in HFGWs, and the angular $\langle\ \rangle$ brackets denote physical quantities averaged over time.

Doing so will lead to detection of relic HFGW, provided the frequency $\omega_g \leq 10^{10} Hz$. Linde et al. (2008) computed an incident relic graviton flux of $N_g \cong 2.89 \times 10^{14} /sec$ at a detector site, as opposed to a graviton flux of $<n>_g \sim 7.872 \times 10^{+21}/sec$ at the onset of the big bang. In addition, a prediction that makes the case for quantum entanglement detectors and Li et al. (2008), where there is a gap between constant value HFGWs gravitons detectable, by the Li-Baker detector, and relic gravity wave HFGW gravitons is given by Eqn (27)
Equations (27), (28) and (29) imply that there is ample room to detect energized gravitons and that astrophysical neutrinos have energies $\leq 10^{12} GeV$, and that for every graviton, there is about $10^5$ neutrinos, according to Baumann *et al*. (2007). This means that, by using

$$N_{g-relic-GW} \approx 2.89 \times 10^{14} << N_{relic-Neutrinos} \approx 10^{20} \qquad (31)$$

and assuming relic gravitons can have frequencies up to $10^2 GHz$ (Brustein, *et al.*, 1995 and 1997), or up to $10^{10} Hz$ (Grishchuk, 2007), it can be concluded that the energy density today in horizon-size gravitational waves is at most a tiny fraction, about $10^4$, of the closure density of the universe. Such waves would correspond to a distribution of gravitons of the Planck epoch that would be produced by a very mild degree of chaos. This requires extraordinary sensitivity to detect. However, just as with photons (but at a much earlier era because of the exceedingly weak interaction of gravitons), gravitons attain a characteristic energy distribution. The nature of quantum gravity and graviton interactions are uncertain and this need not be a blackbody distribution. The high end of graviton energy distribution probably is connected with the difference in mass, i.e., if neutrinos have $\leq 10^{12} GeV$ per individual neutrino, then gravitons would have $\leq 10^{-34} \times 10^{12} GeV$ per graviton; meaning that it would require refined HFGW detectors such as the Li-Baker detector to observe them at all.

If there is a ratio of 1 graviton per ten to the fifth power or so neutrinos, data from refined HFGW detectors should be able to be matched to IceCubeIceCube data.

## Inflation models, neutrino detection and HFGWs

If there is a density increase as given by an inflation $\phi$ occurring because of chaotic inflation

$$\delta_H \approx \frac{H^2}{2\pi\dot{\phi}} \sim \frac{m\phi^2}{5\pi\sqrt{6}}, \tag{32}$$

then the following spectral tilt values can be expected, i.e. for scalar and tensors

$$n_s = 1 - 3 \cdot \left[\frac{\partial V/\partial \phi}{V}\right]^2 + 2\frac{\left[\partial^2 V/\partial \phi^2\right]}{V} \approx .95 \pm .016 \tag{33}$$

and an absolute value of the tensorial spectral tilt which is very small

$$|n_t| = \left[\frac{\partial V/\partial \phi}{V}\right]^2 << 0.1 \tag{34}$$

The smallness of the tensorial spectral tilt is in tandem with the relative faintness of equation (32) above, that so long as $|n_t| \neq 0$, but is still small, giving a realistic situation, especially since $[\partial V/\partial \phi/V]^2 \sim 10^8 \cdot H^2$.

Were if $f \sim 10^{10} Hz$ were used, then one would have no relic gravitational waves at all - or relic gravitons.

## WHY USE SIMPLE CHAOTIC INFLATION: POWER TO INFLATON POTENTIALS

The following discussion is from Linde (2008), but it points out that simple is better in terms of obtaining vigorous inflationary expansion, which ties in nicely with obtaining $\bar{\rho}_{GW}^{(3)}(\tau,\tau_0) \neq 0$. Looking first at a preliminary inflation field expansion, as given by the potential

$$V = V_0 + \alpha\phi + \frac{m^2}{2}\phi^2 + \lambda_n \frac{\phi^{4+n}}{M_{PL}^n} + \frac{\xi}{2} \cdot R \cdot \phi \tag{35}$$

**Case 1.** Looking at the simplest case, where $m^2 \equiv 0 \equiv \lambda_n \equiv \xi$, there is no inflation, since $V_0 + \alpha\phi < 10^{-120}$. Leading to no inflation, since it would make $\alpha$ too small to give density perturbations.

**Case 2.** Optimal where

$$V = V_0 + \frac{m^2}{2}\phi^2. \tag{36}$$

This case allows considering $m^2\phi^2/2 \sim 1 \Leftrightarrow \phi \equiv \phi_{\max imum} \Rightarrow \exp(\phi^2/4) \sim \exp(1/m^2)$ as a maximal inflationary condition.

**Case 3.** Non optimal where

$$V = \frac{m^2}{2}\phi^2 + \lambda_n \frac{\phi^{4+n}}{M_{PL}^n} + \frac{\xi}{2} \cdot R \cdot \phi \tag{37}$$

Maximal inflation would be given if $N < \exp\left[\min\{m^{-2}, \lambda_n^{-2/n}, \xi^{-1}\}\right]$, where the maximal inflationary behavior wanted, if Case 2 is what one works with as a physical model in terms of inflation potentials. Furthermore, as the value get closer and closer to $\phi \equiv o(1)$, i.e. of order 1, where this true even for equation (36); provided $m^2 > \xi, \lambda_n^{2/n}$ and that $V = (m^2/2)\phi^2 + \lambda_n(\phi^{4+n}/M_{PL}^n) + (\xi/2) \cdot R \cdot \phi \xrightarrow{\phi \to o(1)} (m^2/2)\phi_{\max-value}$. In the last stages of inflation, there is convergence to the quadratic behavior of the inflation potential. Furthermore, if any of $m^2 < \xi$ and/or $\lambda_n^{2/n}$ one observes lesser inflation than if $V = (m^2/2)\phi^2$ had been used by an exponential amount. The suppositions and hypothesis written just above are equivalent to $[\partial V/\partial \phi/V]^2 \sim 10^8 \cdot H^2 \neq 0$, especially at the end of the inflationary era.

Now, by optimizing the data sets of the Li-Baker HFGW set, they can be compared with Baret (2006). By noting the case of how chaotic inflation, with $f \sim 10^{10} Hz$ and a quadratic potential helps avoid the disastrous

$$\left[\frac{\partial V/\partial \phi}{V}\right]^2 \sim 10^8 \cdot H^2 = 0 \Leftrightarrow \bar{\rho}_{GW}^{(3)} = 0. \tag{38}$$

That is, one needs

$$\left[\frac{\partial V/\partial \phi}{V}\right]^2 \sim 10^8 \cdot H^2 \neq 0 \Leftrightarrow \bar{\rho}_{GW}^{(3)} \neq 0 \tag{39}$$

Vigorous use of chaotic inflation, with a small but non zero $\bar{\rho}_{GW}^{(3)} \neq 0$, and the largest acceptable value of density fluctuations due to $\delta_H \approx H^2/2\pi\dot{\phi} \sim m\phi^2/5\pi\sqrt{6}$ leads to an increased likelihood of vigorous relic graviton production, i.e. leading to $<n>_g \sim 7.872 \times 10^{+21}/\sec$ at the onset of the big bang. If this large relic graviton production rate is accomplished, partly due to the wormhole model of transferal of energy from a prior universe, then the Baumann *et al*. (2007) value of ten to the fifth relic neutrinos as compared to one relic graviton is not an insurmountable data collection correlation problem. Then, there are enough neutrinos from ICECUBE, and a large number of relic gravitons from the Li-Baker detector to make comparison of the data sets doable. Given $\bar{\rho}_{GW}^{(3)} = 0$, or close to that number, then there would no longer be a possibility of linkage of ICECUBE Neutrinos to any relic Graviton production from the Li-Baker detector. Furthermore, quantum limits should also look at (see; Stephenson, 2009), as a way to enhance the resolution of the neutrino-graviton data match ups, and hope that suitable research and development is initiated in the near future on this topic. This would in addition improve upon the relatively crude estimates of gravitational wave sensitivity reported by Izquierdo (2006) and give credence to the developments given by Baumann, *et al*. (2007) which indicate how to obtain further resolution in early universe HFGW production. It is also argue that additional developments linking HFGW and neutrinos will aid in making an experimental linkage between HFGWs and gravitons and to overcome some of the typical problems traditional detectors have, as noted by Rothman and Boughn (2006).

## CONCLUSIONS

This present article brings up a very subtle point about entanglement which the author will allude to. Giovannini (2008) makes reference to a calculation which he performed in 1993 which suggests that all entropy of the universe from the origins of the big bang, to the present day is due to graviton production. The ratio of the amplitude of HFGW as given in equation (27) with a value of $\sim 10^{-1}$ between the amplitudes of HFGW produced from the big bang, to those obtained at all times as collected at the Li – Baker detector is really due to entanglement indicating a change in spatial geometry of the universe as it is embedded in a higher dimensional structure. I.e. the geometry changed over time If higher dimensional embedding were obtainable, the amplitude value of equation (27) would go to unity, instead of being . $\sim 10^{-1}$ Seen from the perspective of entropy and graviton production, in higher dimensions, if a physical scientist and/ or engineer had a sufficient dimensional reference point to look at the universe, the ratio of amplitude of gravitational waves at the origin of the big bang, and those as seen up to the present would be the same, not differentiated as indicated in the reference point assumed in setting up equation(27). Entanglement is really a measurement of dimensional warping of space, a point which the author will clarify in a future article.

Entanglement indicating an almost instantaneous transfer of information in three dimensional space, can be seen as representing in higher dimensional geometry that points which in three space are far apart are in fact neighbors, and actually close together in higher dimensions. If this rule of thumb is applied to with respect to graviton production, it means that physical scientists are studying the very origins of space time nucleation, and not just a measurement of HFGW in the Li-Baker detector.

In this paper, a linkage is established between how to get the FFT of a second-order tensorial contribution of HFGWs and relic HFGWs from the time of the big bang. As noted in the discussion about equation (27), there are significant difficulties in separating out relic HFGW inputs from the big bang from HFGWs that the Li-Baker detector would get from HFGWs for all times in the universe's evolution from the big bang. This paper also suggests explanations for the relationship between HFGWs, as detected by the Li-Baker HFGW detector, and gravitons as "particles," which is difficult, as noted by Rothman and Boughn (2006).

The key motivations of this paper were to aid in making an experimental linkage between HFGWs and gravitons and to overcome some of the typical problems traditional detectors have, as noted by Rothman and Boughn (2006). The Li-Baker HFGW detector (Li *et al*., 2008) is shown to suggest a way to formally make the HFGW and graviton linkage explicit in data sets. Furthermore, in refinement of procedures for obtaining better HFGW relic data, this paper points out that Yeo et al. (2006) suggests an entanglement entropy-based detector concept for adding resolution to obtain more differentiation between HFGW contributions from the big bang and those HFGW from all times which are detected at the site of the Li-Baker detector. to the already sensitive resolution implied by equation (27). In addition, the analysis leading to Table 1 adds evidence for the existence of gravitons as a measurable physical entity, and suggests detector technology that overcomes the physical limits Rothman and Boughn (2006) postulated for detectors: that they would have to have the diameter of Jupiter in order to detect one graviton a year, a physical measurement absurdity.

In Beckwith (2008), it was suggested how a wormhole construction from a prior universe to our present universe could take place. Support for the wormhole hypothesis of a prior universe contributing a vacuum energy to our present universe is the abrupt rise in temperatures (as given in Table 1), allowing for a relic graviton burst. If a wormhole contributed to early-universe vacuum nucleation, physicists might be able to observe the evolution of early-universe cosmology in the context of the mega-structures larger than the observed universe suggested by Erickcek *et al*. (2008). If these mega-structures exist, discussions of parallel universes, as discussed by Tegmark (2003), are likely necessary. If parallel universes can be ascertained by HFGW data obtained by the Li-Baker detector, there will be exciting times in future cosmology.

Even if the existence of a structure larger than our universe cannot be inferred, being able to identify an experimentally graceful exit from inflation (based on HFGW data from the Li-Baker detector) would be a huge improvement in our current astrophysical understanding of how the universe evolved from the origins of the big bang itself. How could the expansion rate of the big bang slow down, from increasing acceleration to a slowing-down universe expansion? The Li-Baker detector could help us investigate whether or not the graceful exit from inflation actually occurred, if it was a relatively sharp change from an increase in expansion to a decrease in the rate of expansion, or if not, if the process constitutes a phase-order transition in the first place.

Then, there is the question of whether or not the total entropy of the universe stabilized after a sharp increase. Currently, Roos (2003) marks the main burst of entropy increase as due to reheating, which is significantly after the big bang, and models it in terms of GUT arguments. He also argues that this sudden increase in entropy at a time significantly after the big bang violates his expectation that after the big bang, the time derivative of entropy is zero. That is, entropy after the big bang stabilizes. The question is then, "Does entropy, abruptly taper off, increase, or change in other ways?" An HFGW detector may be our only way to answer this question.

Finally let us review the tie in with Gravitons and relic neutrinos. Now, by optimizing the data sets of the Li-Baker HFGW set, they can be compared with Baret (2006). By noting the case of how chaotic inflation, with $f \sim 10^{10} Hz$ and a quadratic potential helps avoid the disastrous

$$\left[\frac{\partial V/\partial \phi}{V}\right]^2 \sim 10^8 \cdot H^2 = 0 \Leftrightarrow \bar{\rho}_{GW}^{(3)} = 0. \tag{40}$$

That is, one needs

$$\left[\frac{\partial V/\partial \phi}{V}\right]^2 \sim 10^8 \cdot H^2 \neq 0 \Leftrightarrow \bar{\rho}_{GW}^{(3)} \neq 0 \tag{41}$$

Vigorous use of chaotic inflation, with a small but non zero $\bar{\rho}_{GW}^{(3)} \neq 0$, and the largest acceptable value of density fluctuations due to $\delta_H \approx H^2/2\pi\dot{\phi} \sim m\phi^2/5\pi\sqrt{6}$ leads to an increased likelihood of vigorous relic graviton production, i.e. leading to $<n>_g \sim 7.872 \times 10^{+21}/\sec$ at the onset of the big bang. If this large relic graviton production rate is accomplished, partly due to the wormhole model of transferal of energy from a prior universe, then the Baumann *et al*. (2007) value of ten to the fifth relic neutrinos as compared to one relic graviton is not an insurmountable data collection correlation problem. Then, there are enough neutrinos from ICECUBE, and a large number of relic gravitons from the Li-Baker detector to make comparison of the data sets doable. Given $\bar{\rho}_{GW}^{(3)} = 0$, or close to that number, then there would no longer be a possibility of linkage of ICECUBE Neutrinos to any relic Graviton production from the Li-Baker detector. Furthermore, quantum limits should also look at (see; Stephenson, 2009), as a way to enhance the resolution of the neutrino-graviton data match ups, and hope that suitable research and development is initiated in the near future on this topic. This would in addition improve upon the relatively crude estimates of gravitational wave sensitivity reported by Izquierdo (2006) and give credence to the developments given by Baumann, *et al*. (2007) which indicate how to obtain further resolution in early universe HFGW production. It is also argue that additional developments linking HFGW and neutrinos will aid in making an experimental linkage between HFGWs and gravitons and to overcome some of the typical problems traditional detectors have, as noted by Rothman and Boughn (2006).

## NOMENCLATURE

$a(t)$ = scale factor $\left[ \equiv (1+z_{red-shift})^{-1} = \lambda_{rest}/\lambda_{observed} \right]$

$c$ = speed of light (m/s)

$g_{uv}$ = metric tensor $[\equiv \eta_{uv} + h_{uv}]$

$G$ = gravitational constants $\left[ 6.673 \times 10^{-11} m^3/kgs^2 \right]$

$\hbar$ = reduced Planck constant $\left[ \cong 1.054 \times 10^{-34} J \cdot \sec \right]$

$J$ = energy (J)

$k$ = $(2\pi/\lambda)$ – part of wave vector

$L$ = length (m)

$\tau$ = conformal time $\left[ \equiv \int dt/a(t) \right]$

$t_P$ = Planck time

$T$ = temperature (K)

$T^{uv}$ = GR – energy-stress-tensor

$T^{00}$ = energy-density

$dS^2$ = arc length, squared of general relativity $\left[ \equiv g_{uv} x^u x^v \right]$

$dS^2$ = arc length, squared of general relativity $\left[ \equiv g_{uv} x^u x^v \right]$

$\hat{h}$ = scalar rendition of Fourier transform of variation of metric Tensor $g_{uv}$ from flat space metric $\eta_{uv}$

$h_{uv}$ = gravitational wave contribution to metric distance from the observer

$\lambda_{observed}$ = wavelength of cosmological objects observed on the Earth's surface

$\lambda_{rest}$ = wavelength of cosmological objects in their rest frame about themselves, when far from the Earth's surface

$m_{graviton}$ = $\sim 10^{-60}$ (kg)

$z_{red-shift}$ = red shift $\left[ \equiv (\lambda_{observed} - \lambda_{rest})/\lambda_{rest} \right]$

## ACKNOWLEDGEMENTS

The author wishes to thank Dr. Eric Davis and Dr. Robert Baker for stimulating his interest in gravitation, and gravitational waves. Furthermore, I appreciate the assistance of Amara Angelica, who converted a previously hard-to-understand document into a more readable version.